%% file: main.tex
\documentclass[]{valhalla}
\input{math_commands.tex}

\input{styles/epibench_macros.tex}

\paperlogo{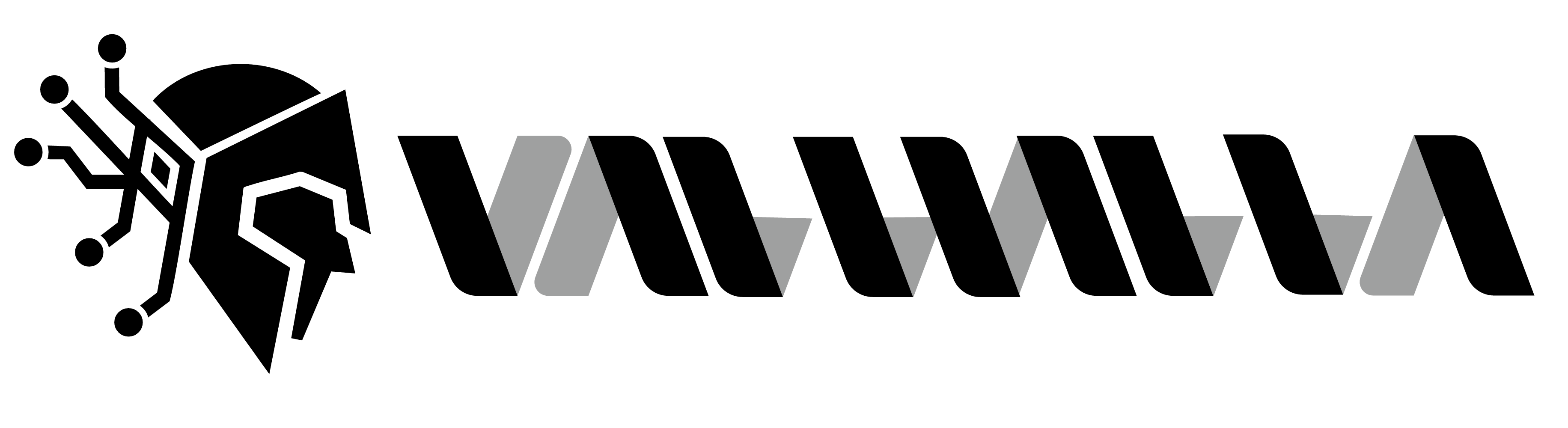}
\papertagline{[Paper tagline]}

\title{Leveraging secondary-structure information for accurate nucleic acid structure prediction with OFoldNA}
\author[*]{Valhalla Team}

\abstract{Recent advances in biomolecular structure prediction have enabled accurate modelling of increasingly complex molecular systems. However, nucleic acid structure prediction remains challenging because of conformational flexibility and the limited availability of high-quality 3D structural data. Secondary structure (SS) provides a more readily available layer of structural information that captures base-pairing relationships and folding topology. Here we present OFoldNA, an all-atom diffusion model that incorporates SS information into nucleic acid folding and protein--nucleic acid co-folding. Without external SS information, OFoldNA achieved leading performance on FoldBench for both nucleic acid monomer folding and protein--nucleic acid co-folding, with particularly strong performance on DNA monomers and protein--DNA interfaces involving longer nucleic acid chains. When accurate base-pairing information was provided, OFoldNA-SS2TS further improved both folding and co-folding accuracy, while partial SS information also yielded consistent gains. The same auxiliary branch can also be used for RNA SS prediction as OFoldNA-SS, which achieved the best out-of-distribution performance on CHANRG. Together, these results show that intermediate structural information such as nucleic acid SS can be leveraged to improve all-atom 3D modelling, providing a general direction for incorporating complementary structural modalities into molecular structure prediction and design.}

\checkdata[Code]{\url{https://github.com/OTeam-AI4S/OFoldNA}}

\begin{document}
\maketitle

\begingroup
\renewcommand{\thefootnote}{\fnsymbol{footnote}}
\footnotetext[1]{Full author list in Contributions}
\endgroup

\input{sections/01_introduction.tex}
\input{sections/04_results.tex}
\input{sections/05_discussion.tex}
\input{sections/03_methods.tex}

\section*{Contributions}
Odin Zhang conceived the study.
Yuntao Yu and Zhiyuan Chen designed and implemented the network.
Yuntao Yu, Heqin Zhu, Zhiyuan Chen, Kejun Yin, Jianmin Wang, Weibo Zhao,
Qinghan Wang, Ziyi You, Gang Du and Lijun Lang designed and conducted
the experiments and analyzed the results.
Yuntao Yu drafted the manuscript.
All authors discussed the results and revised the manuscript.
\bibliographystyle{valhalla_references}
\bibliography{references}

\clearpage
\appendix
\setcounter{table}{0}
\renewcommand{\thetable}{\arabic{table}}
\renewcommand{\theHtable}{supplementary.\arabic{table}}
\captionsetup[table]{
    name=Supplementary Table,
    labelsep=period,
    format=plain,
    labelfont={sf,bf}
}
\EpibenchAppendixHeadingSpacing
\input{appendix/01_appendix.tex}

\end{document}

%% file: math_commands.tex
\usepackage{amsmath,amsfonts,bm}

\def\eqref#1{equation~\ref{#1}}

\def\1{\bm{1}}

\DeclareMathAlphabet{\mathsfit}{\encodingdefault}{\sfdefault}{m}{sl}
\SetMathAlphabet{\mathsfit}{bold}{\encodingdefault}{\sfdefault}{bx}{n}



%% file: styles/epibench_macros.tex
\usepackage{iftex}
\usepackage{url}
\usepackage{booktabs}
\usepackage{graphicx}
\usepackage{float}
\usepackage{xcolor}
\usepackage{colortbl}
\usepackage{makecell}
\usepackage{pifont}
\usepackage{wrapfig}
\usepackage{hyperref}
\usepackage{array}
\usepackage{multirow}
\usepackage{amsmath}
\usepackage{amssymb}
\usepackage{amsfonts}
\usepackage{nicefrac}
\usepackage{microtype}
\usepackage{longtable}
\usepackage{enumitem}
\usepackage{subcaption}
\usepackage{cleveref}
\usepackage[toc,page,header]{appendix}
\usepackage{minitoc}
\usepackage{tcolorbox}
\tcbuselibrary{most}

\makeatletter
\newcommand{\EpibenchAppendixHeadingSpacing}{%
    \renewcommand{\section}{\@startsection{section}{1}{\z@}{-0.8ex plus -0.2ex minus -.1ex}{0.25ex plus .08ex}{\large\sffamily\bfseries\color{valhallaprimary}\raggedright}}%
    \renewcommand{\subsection}{\@startsection{subsection}{2}{\z@}{-0.55ex plus -0.16ex minus -.08ex}{0.16ex plus .05ex}{\normalsize\sffamily\bfseries\color{valhallaprimary}\raggedright}}%
    \renewcommand{\subsubsection}{\@startsection{subsubsection}{3}{\z@}{-0.18ex plus -0.06ex minus -.04ex}{0.05ex plus .02ex}{\normalsize\sffamily\bfseries\color{valhallaprimary}\raggedright}}%
}
\newcommand{\EpibenchNeedspace}[1]{%
    \par\begingroup
    \dimen@=#1\relax
    \dimen@ii=\pagegoal
    \advance\dimen@ii by -\pagetotal
    \ifdim\dimen@>\dimen@ii
        \pagebreak
    \fi
    \endgroup
}
\makeatother

\newcounter{epialgorithm}

\definecolor{ReasoningBand}{HTML}{F1EAFE}
\definecolor{NonreasoningBand}{HTML}{F7F3EE}
\newsavebox{\exampleboxcontent}
\newsavebox{\systempromptboxcontent}
\newsavebox{\caseboxcontent}
\newlength{\exampleboxwidth}
\newlength{\systempromptboxwidth}
\newlength{\caseboxwidth}

%% file: sections/01_introduction.tex
\section{Introduction}

Nucleic acids lie at the core of biological information storage, transmission and regulation~\cite{crick1970central}. DNA stores hereditary information, while its structural properties also contribute to gene regulation and genome maintenance~\cite{wang2023dynamic,wulfridge2024intertwining,petermann2022sources}. RNA not only conveys genetic information but also performs diverse regulatory, structural and catalytic functions~\cite{cao2024identification,ganser2019roles,dykstra2022engineering}. These roles are determined not by sequence alone: like proteins, many nucleic acids adopt specific secondary and tertiary structures and sample distinct conformational states that are integral to their biological activity. Determining the three-dimensional (3D) structures of nucleic acids is therefore critical to understanding their mechanisms of action~\cite{spitale2023probing}, characterizing interactions with proteins and other biomolecules~\cite{ganser2019roles,ducoli2026bridging}, and establishing a structural basis for the development of nucleic acid–targeted therapeutics~\cite{childs2022targeting}. Despite their biological importance, the conformational flexibility of nucleic acids complicates the determination of the experimental structure~\cite{wang2026integrated}. Consequently, nucleic acid structures remain substantially underrepresented in the Protein Data Bank (PDB). As of August 2026, proteins, peptides and viruses accounted for 85.8\% of the PDB, whereas protein--nucleic acid complexes and entries containing only nucleic acids accounted for just 6.5\% and 2.0\%~\cite{berman2000pdb}.

Secondary structure (SS) provides a complementary and substantially more accessible source of structural information~\cite{zhang2022advances}. Base-pairing relationships define much of the folding topology of nucleic acids, placing SS at an intermediate level between sequence and atomic structure. Importantly, SS information can often be obtained at scales that are difficult to achieve with high-resolution 3D structure determination. For RNA, high-throughput structure-probing methods such as SHAPE-MaP~\cite{siegfried2014rna}, icSHAPE~\cite{spitale2015structural}, and DMS-MaPseq~\cite{zubradt2017dms} generate nucleotide-resolution reactivity profiles from individual RNAs to transcriptome-wide measurements and can be used to inform SS inference. For DNA, base-pairing patterns in canonical duplexes can often be determined directly from strand complementarity. More complex DNA structures, including G-quadruplexes~\cite{chambers2015high,bedrat2016re}, i-motifs~\cite{zeraati2018motif,yu2024seeker}, and cruciforms~\cite{kouzine2017permanganate,brazda2016palindrome}, can be identified experimentally or predicted using specialized computational methods. Given the scarcity of high-quality 3D structural data, incorporating SS information as a prior on folding topology offers a practical strategy for improving the accuracy of nucleic acid 3D structure modeling~\cite{li2023integrating,wang2023trrosettarna,tarafder2026rnabpflow}.

The use of SS information in 3D structure prediction has been explored most extensively for RNA monomers. Conventional RNA structure prediction methods, such as RNAComposer~\cite{popenda2012automated}, 3dRNA v2~\cite{wang20193drna}, FARFAR2~\cite{watkins2020farfar2}, and SimRNA~\cite{boniecki2016simrna} 
reconstruct RNA structures through template or fragment assembly and conformational sampling, often with SS information used to constrain the folding process. More recent deep-learning methods, including DRfold~\cite{li2023integrating}, trRosettaRNA~\cite{wang2023trrosettarna}, NuFold~\cite{kagaya2025nufold}, and RNAbpFlow~\cite{tarafder2026rnabpflow}, incorporate predicted or experimentally derived SS information more directly into structure prediction, while RhoFold+~\cite{shen2024accurate} introduces SS supervision during training. Despite these advances, the use of SS information has remained largely confined to RNA monomer prediction. Progress in DNA structure prediction has generally lagged behind RNA. Available methods are relatively few and often designed for specific modelling tasks. 3dDNA~\cite{zhang20223ddna} extends SS-guided template assembly to DNA, whereas DNAfold2~\cite{wang20253d} uses coarse-grained conformational sampling to predict branched DNA structures. In parallel, general purpose all-atom models, including RoseTTAFold2NA~\cite{baek2024accurate}, AlphaFold 3~\cite{abramson2024accurate}, Boltz-1~\cite{wohlwend2025boltz}, Chai-1~\cite{chai2024chai}, and Protenix v1~\cite{protenix2026protenix}, have broadened structure prediction to multicomponent complexes containing proteins, nucleic acids, and other biomolecules. However, nucleic acids remain a comparatively data-scarce modality for 3D structure prediction, making it difficult to rely on atomic structural supervision alone. Meanwhile, substantially more abundant SS information is available for nucleic acids but has not been systematically leveraged by general all-atom models, leaving a gap between scalable nucleic acid SS information and all-atom 3D structure prediction.

In practice, however, nucleic acid SS information is heterogeneous in both its source and completeness. Experimental structure-probing measurements often provide nucleotide-level reactivity profiles rather than explicit base-pairing relationships, whereas sequence analysis and computational prediction can provide candidate pairing patterns with varying levels of confidence. After being translated into pairing information, the resulting structural priors may therefore range from complete maps to partially resolved or uncertain subsets. General structure prediction models should therefore not assume a single fixed form of SS input. Instead, such models should represent pairing information in a common format and operate across different levels of SS availability, predicting from sequence when no SS is provided while progressively exploiting partial or complete pairing information when available.

Here we introduce OFoldNA, a unified all-atom diffusion model for nucleic acid monomer folding and protein--nucleic acid co-folding that is designed to leverage nucleic acid SS information at varying levels of availability. OFoldNA embeds base-pairing information through a dedicated SS embedding module and learns these relationships internally through an auxiliary SS prediction objective. Random masking of pairing maps during training enables the model to operate under absent, partial, and complete SS conditions, while optional classifier-free guidance modulates the influence of supplied SS information during sampling without retraining. On FoldBench, OFoldNA achieved leading performance in nucleic acid monomer folding and protein--nucleic acid co-folding, with particularly strong gains for DNA and protein--DNA interfaces involving longer nucleic acid chains. The auxiliary prediction branch, OFoldNA-SS, also showed a strong out-of-distribution generalization in RNA SS prediction. When accurate pairing information was supplied, OFoldNA-SS2TS further improved all-atom structure prediction, with gains emerging even from partial SS information. All these results establish OFoldNA as an all-atom framework that connects nucleic acid sequence, scalable SS information, and 3D structure prediction.

%% file: sections/04_results.tex
\section{Results}

\subsection{Overview of OFoldNA framework}

\begin{figure}[!t]
    \centering
    \includegraphics[width=\linewidth]{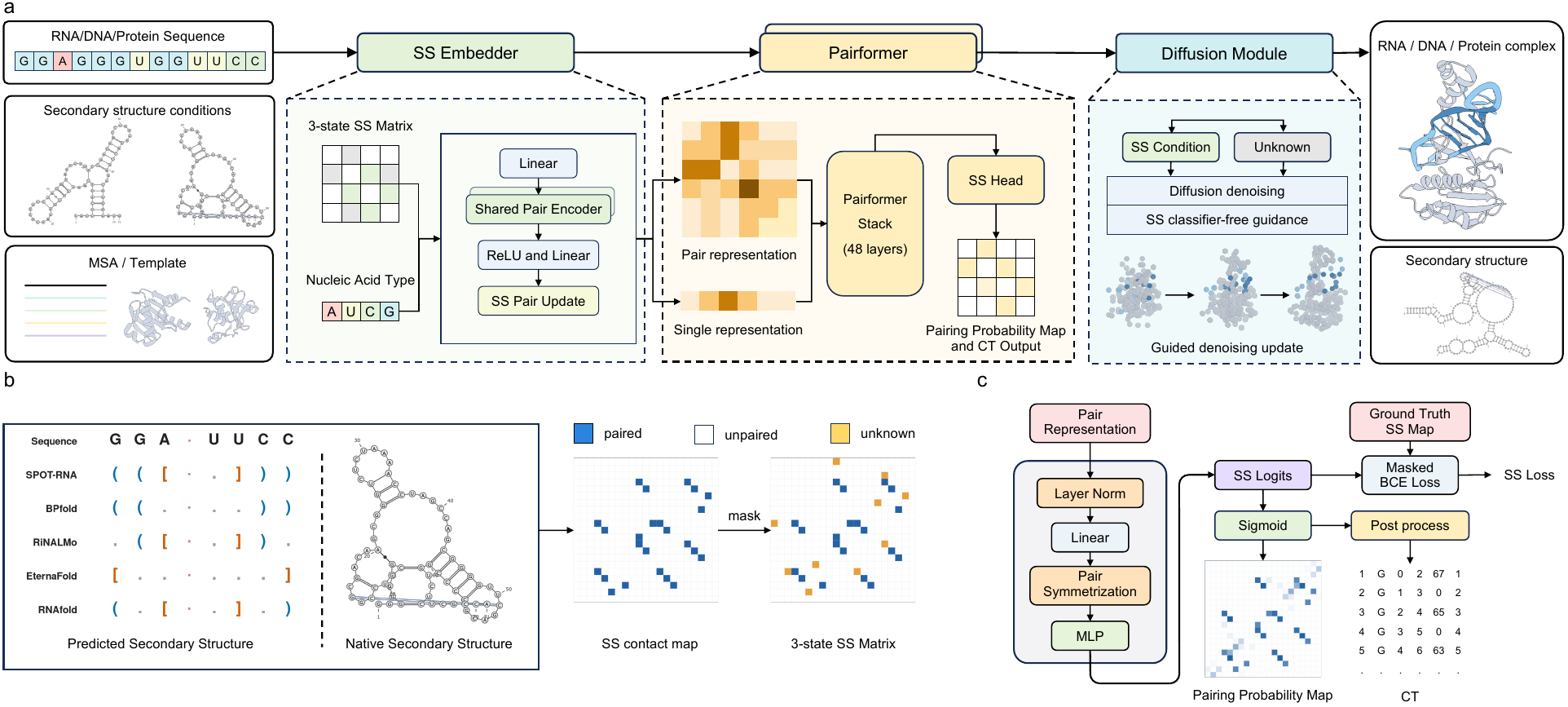}
    \caption{\textbf{Overview of the OFoldNA pipeline.} \textbf{a,} Overall model architecture. The SS Embedder encodes a three-state pairing map (paired, unpaired and unknown) as an update to the pair representation, which is injected into the Pairformer trunk. The resulting Pairformer representations are used for SS prediction and diffusion based all-atom structure generation. During sampling, SS conditioned classifier-free guidance combines denoising predictions from a branch conditioned on the supplied SS and an all-unknown reference branch. \textbf{b,} Masking augmentation for SS conditioning. During training, either the entire pairing map or a subset of its entries is randomly masked as unknown, exposing the model to partially observed and fully unknown SS conditions. \textbf{c,} SS prediction head. The SS Head predicts base-pairing probabilities from the Pairformer pair representation and is trained with a masked binary cross-entropy loss that excludes unknown entries. The predicted probability map can be converted into an SS in connectivity table (CT) format during post-processing.}
    \label{fig:model-overview}
\end{figure}

The overall architecture of OFoldNA is illustrated in Fig.~\ref{fig:model-overview}a. A dedicated SS embedding module encodes a three-state pairing map (paired, unpaired and unknown) into pair representations. To accommodate practical settings in which SS information is only partially known or entirely unavailable, we randomly mask either the full map or a subset of its entries as unknown during training. This exposes the model to different levels of SS availability and enables prediction under partial or absent external SS conditions (Fig.~\ref{fig:model-overview}b). In addition, an SS prediction head is applied to the Pairformer pair representations and trained with a masked binary cross-entropy loss to provide auxiliary supervision for learning base-pairing relationships (Fig.~\ref{fig:model-overview}c). This auxiliary objective encourages the model to encode base-pairing knowledge directly in its internal representations, allowing such structural information to be retained even when no external pairing map is provided. Without external SS information, OFoldNA performs all-atom structure prediction directly from the standard molecular input. When a complete or partial pairing map is provided, we refer to the corresponding SS-conditioned structure prediction setting as OFoldNA-SS2TS. The supplied pairing information is incorporated into the pair representations through the SS embedding module to guide 3D structure generation. During diffusion sampling, OFoldNA-SS2TS can further strengthen the influence of the supplied SS information using classifier-free guidance (CFG). CFG combines denoising predictions from an SS-conditioned branch and an all-unknown reference branch, allowing the strength of SS conditioning to be adjusted at inference time without retraining. The auxiliary SS prediction head can also be used independently as a standalone SS predictor, termed OFoldNA-SS. OFoldNA-SS outputs a base-pairing probability map that can be converted into an SS in connectivity table (CT) format.

\subsection{OFoldNA enables accurate all-atom folding of nucleic acid monomers}

\begin{figure}[!t]
    \centering
    \includegraphics[width=\linewidth]{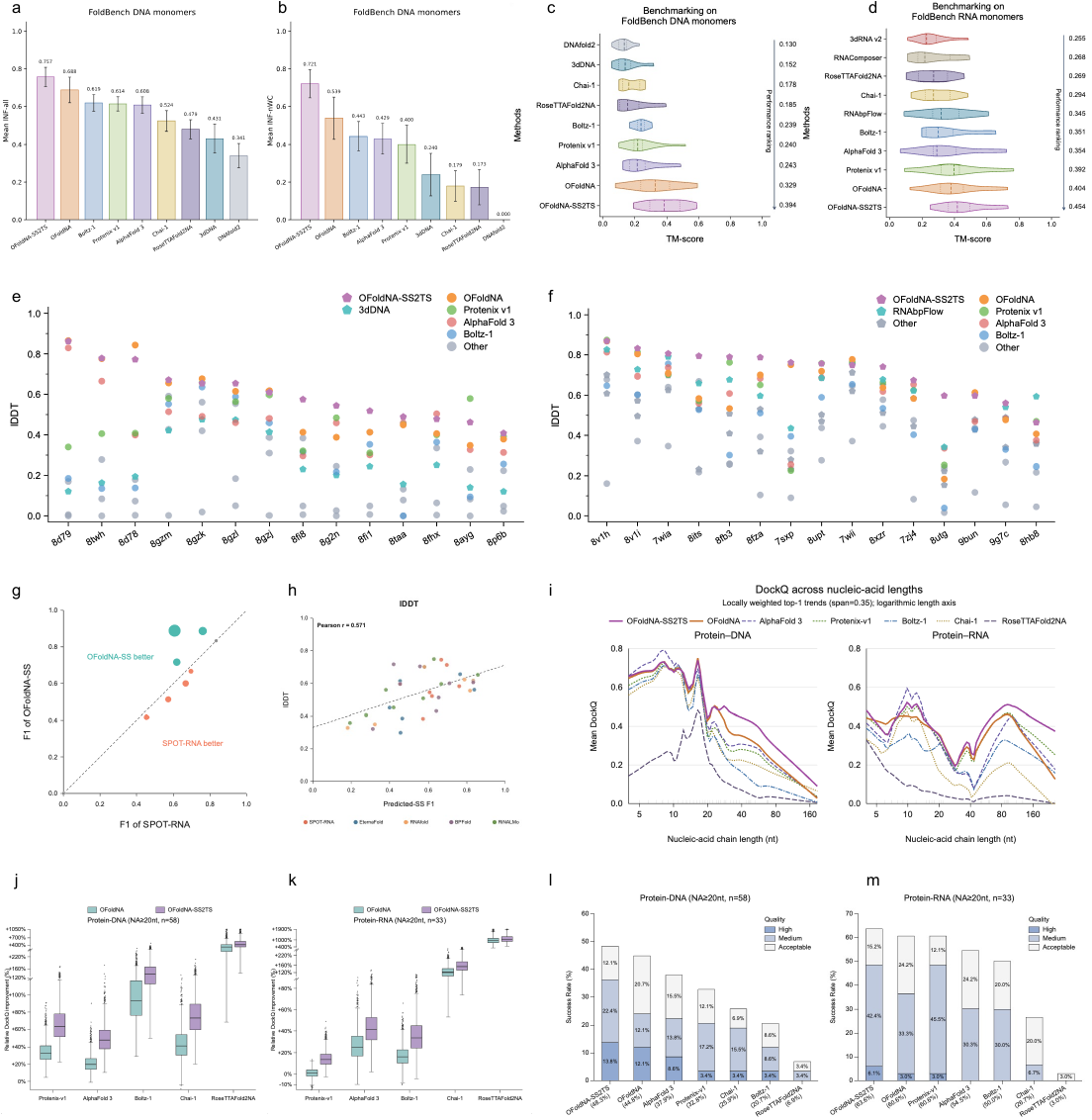}
    \caption{\textbf{Benchmarking OFoldNA for nucleic acid monomer folding and protein--nucleic acid co-folding on FoldBench.} \textbf{a,b,} Mean INF across 14 DNA monomer targets for all interactions (INF-all; \textbf{a}) and non-Watson--Crick pairs (INF-nWC; \textbf{b}). \textbf{c,d,} TM-score distributions across 14 DNA (\textbf{c}) and 15 RNA (\textbf{d}) monomer targets. \textbf{e,f,} Target-level lDDT values for the same DNA (\textbf{e}) and RNA (\textbf{f}) targets. Pentagons indicate methods supplied with reference-derived SS, circles indicate methods without such information. \textbf{g,} Head-to-head comparison of base-pair $F_1$ between OFoldNA-SS and SPOT-RNA. \textbf{h,} Pairing-map $F_1$ versus lDDT of the corresponding rank-1 OFoldNA-SS2TS predictions for five external SS predictors. The dashed line indicates the linear fit (Pearson's $r=0.571$). \textbf{i,} Locally weighted trends in DockQ as a function of nucleic acid chain length for protein--DNA and protein--RNA interfaces. Curves were estimated with a span of 0.35. \textbf{j,k,} Paired bootstrap distributions of the relative improvement in mean DockQ for OFoldNA and OFoldNA-SS2TS over each baseline on protein--DNA (\textbf{j}) and protein--RNA (\textbf{k}) subsets containing nucleic acid chains of at least 20 nucleotides. \textbf{l,m,} DockQ success rates for protein--DNA (\textbf{l}) and protein--RNA (\textbf{m}) interfaces under the same length filter.}
    \label{fig:foldbench-benchmark}
\end{figure}

To evaluate nucleic acid monomer prediction, we compared OFoldNA with general biomolecular structure predictors and nucleic acid-specific methods on the nucleic acid monomer subsets of FoldBench~\cite{xu2025foldbench}. All methods received the nucleic acid sequence as input. Ions and small molecules present in the reference structures were excluded, whereas nucleotide modifications were provided to methods that support them. We evaluated global structure accuracy using GDT-TS, TM-score and RMSD, local accuracy using lDDT, and the recovery of base-pairing and stacking interactions using interaction network fidelity (INF). Whereas coordinate-based metrics measure global and local structural similarity, INF evaluates the recovery of interaction networks that define nucleic acid topology and local geometry. Among methods without external SS information, OFoldNA ranked first in 12 of the 16 DNA- and RNA-level metric comparisons (Table~\ref{tab:dna-rna-monomer-rank1}). On DNA, OFoldNA achieved a mean TM-score of 0.329, approximately 35\% higher than AF3, and a mean lDDT of 0.560, representing a 20.9\% improvement over Protenix-v1. At the interaction network level, OFoldNA achieved an INF-all of 0.688 and an INF-nWC of 0.539, exceeding the corresponding Boltz-1 values by 11.2\% and 21.6\%, respectively (Fig.~\ref{fig:foldbench-benchmark}a,b). On RNA, OFoldNA achieved the highest mean TM-score among methods without external SS information (0.404) and remained competitive even with RNAbpFlow, which was supplied with base-pairing maps extracted from the reference structures. OFoldNA achieved a comparable mean lDDT (0.623 versus 0.636), higher INF-all, INF-stack and INF-nWC, and a slightly lower INF-WC. Using TM-score $>0.45$ as the criterion for correct global-fold recovery~\cite{gong2019rna}, OFoldNA recovered correct global folds for 3 of 14 DNA targets (21.4\%), compared with 1 of 14 (7.1\%) for Protenix-v1 (Fig.~\ref{fig:foldbench-benchmark}c). On RNA, OFoldNA recovered correct global folds for 4 of 15 targets (26.7\%), compared with 3 of 15 (20.0\%) for the best baseline under this criterion (Fig.~\ref{fig:foldbench-benchmark}d).

To evaluate whether OFoldNA could further benefit from external SS information, we supplied base-pairing maps extracted from the reference structures to OFoldNA-SS2TS. Relative to OFoldNA, this setting improved mean TM-score by 19.6\%, lDDT by 8.0\%, INF-all by 10.1\% and INF-nWC by 33.8\% on DNA. The corresponding gains on RNA were 12.2\%, 15.3\%, 9.7\% and 37.3\% (Table~\ref{tab:dna-rna-monomer-rank1}). The global-fold success rate further increased from 21.4\% (3 of 14) to 42.9\% (6 of 14) for DNA and from 26.7\% (4 of 15) to 46.7\% (7 of 15) for RNA (Fig.~\ref{fig:foldbench-benchmark}c,d). In target-wise comparisons, OFoldNA-SS2TS outperformed all baseline methods in lDDT on 11 of 14 DNA targets (78.6\%; Fig.~\ref{fig:foldbench-benchmark}e) and 12 of 15 RNA targets (80.0\%; Fig.~\ref{fig:foldbench-benchmark}f). Thus, reference-derived SS improved both structural accuracy and interaction network recovery. To assess how pairing-map quality related to OFoldNA-SS2TS performance, we generated pairing maps for the 15 RNA monomer targets using five external SS predictors~\cite{singh2019spot-rna,zhu2025bpfold,penic2025rinalmo,wayment2022eternafold,lorenz2011viennarnafold} and supplied the resulting maps to OFoldNA-SS2TS (Supplementary Table~\ref{tab:rna-monomer-predicted-ss-rank1}). Across these targets, OFoldNA-SS achieved the highest base-pair $F_1$ of 0.368, compared with 0.347 for SPOT-RNA (Fig.~\ref{fig:foldbench-benchmark}g) and 0.265--0.320 for the other external predictors. Across the 39 method--target combinations with non-zero base-pair $F_1$ from the five external predictors, pairing-map $F_1$ was positively correlated with the lDDT of the corresponding OFoldNA-SS2TS predictions (Pearson correlation coefficient, 0.571; Fig.~\ref{fig:foldbench-benchmark}h).

\input{tables/monomer_benchmark.tex}

\subsection{OFoldNA improves protein–nucleic acid co-folding, particularly for longer nucleic acids}
We evaluated protein--nucleic acid co-folding on the protein--DNA and protein--RNA subsets of FoldBench (Table~\ref{tab:protein-na-interface-rank1-all-lengths}). Without external SS information, OFoldNA achieved the strongest overall performance on protein-DNA targets, with  the highest mean DockQ (0.604), lowest iRMSD (4.12\,\AA{}) and lRMSD (9.95\,\AA{}), and highest lDDT (0.851) among the methods compared. On protein--RNA targets, OFoldNA achieved the highest mean DockQ (0.400) and lowest lRMSD (16.45\,\AA{}), while its iRMSD and lDDT remained close to the best baseline values. On interfaces involving shorter nucleic acid chains, current state-of-the-art predictors already achieved relatively high accuracy. For example, on protein--DNA interfaces with nucleic acid chains shorter than 20 nucleotides, AF3 achieved a success rate of 85.8\% and a mean DockQ of 0.675; the corresponding values on protein--RNA interfaces were 73.0\% and 0.425. However, performance generally declined as nucleic acid chain length increased (Fig.~\ref{fig:foldbench-benchmark}i). We therefore focused on the more challenging subset containing nucleic acid chains of at least 20 nucleotides, comprising 58 protein--DNA and 33 protein--RNA interfaces (Supplementary Table~\ref{tab:protein-na-interface-rank1}). On the resulting protein--DNA subset, OFoldNA achieved a mean DockQ of 0.286, outperforming AF3 (0.238) and Protenix-v1 (0.215) by 20.2\% and 33.1\%. Paired bootstrap analysis yielded similar median improvements of 19.7\% and 32.9\% (Fig.~\ref{fig:foldbench-benchmark}j). OFoldNA reached a success rate of 44.8\%, 6.9 percentage points higher than AF3, and produced medium- or high-quality predictions for 24.1\% of interfaces (Fig.~\ref{fig:foldbench-benchmark}l). On the protein--RNA subset, OFoldNA performed comparably to Protenix-v1, achieving a mean DockQ of 0.362 versus 0.359 and identical success rates of 60.6\% (Fig.~\ref{fig:foldbench-benchmark}k,m).

\input{tables/interface_full.tex}

Providing SS information extracted from the reference complexes further improved performance on the same length-filtered subsets. On the protein--DNA interfaces, OFoldNA-SS2TS increased mean DockQ by 22.9\% and reduced lRMSD by 27.1\% relative to OFoldNA. Its median paired-bootstrap DockQ improvements over AF3 and Protenix-v1 reached 47.4\% and 63.3\% (Fig.~\ref{fig:foldbench-benchmark}j). The success rate increased to 48.3\%, while the fraction of medium- or high-quality predictions rose from 24.1\% to 36.2\%; high-quality predictions alone increased to 13.8\%, compared with 8.6\% for AF3 (Fig.~\ref{fig:foldbench-benchmark}l). A similar trend was observed on protein--RNA interfaces. Relative to OFoldNA, OFoldNA-SS2TS increased mean DockQ by 12.9\%, reduced iRMSD by 23.5\%, and increased the fraction of medium- or high- quality predictions from 36.4\% to 48.5\%, with an overall success rate of 63.6\% (Fig.~\ref{fig:foldbench-benchmark}m). The corresponding median paired-bootstrap DockQ improvements over AF3 and Protenix-v1 were 41.2\% and 13.6\% (Fig.~\ref{fig:foldbench-benchmark}k). These results demonstrate that OFoldNA is competitive on protein--nucleic acid co-folding, particularly for protein--DNA interfaces, and that explicit SS conditioning yields broader gains on the more challenging interfaces involving longer nucleic acid chains.

\subsection{OFoldNA recovers challenging nucleic acid folds and multicomponent assemblies}
\begin{figure}[!t]
    \centering
    \includegraphics[width=\linewidth]{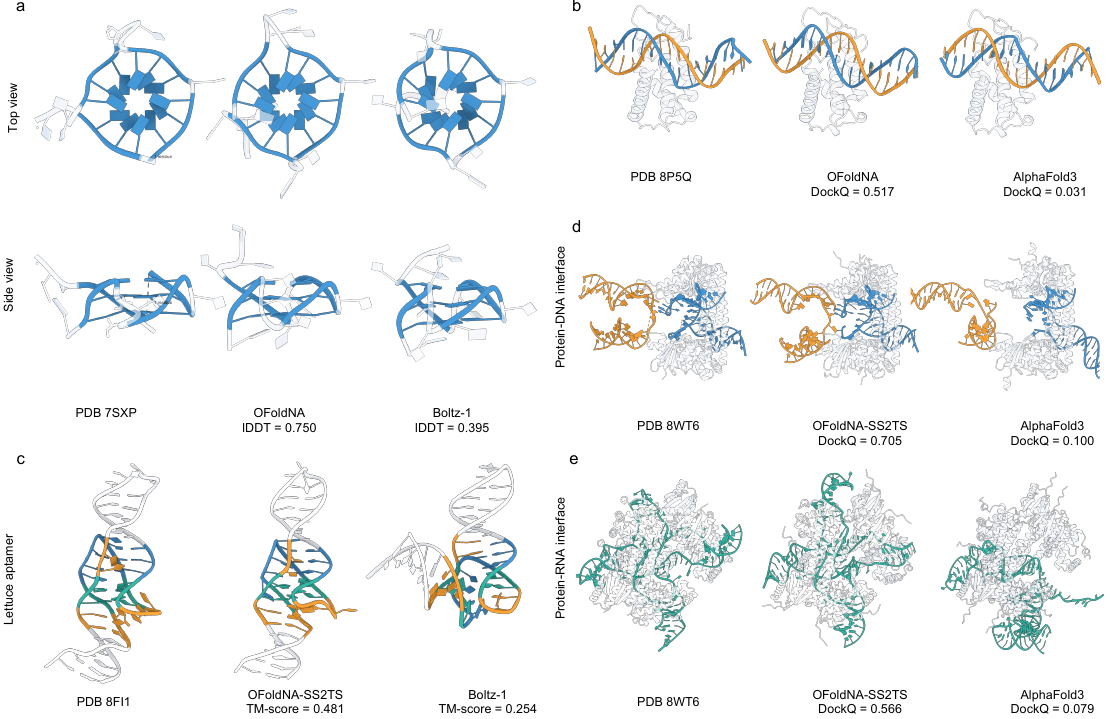}
    \caption{\textbf{Representative structure predictions from OFoldNA and OFoldNA-SS2TS.} \textbf{a,} Experimental structure of the 22-nucleotide G8U variant of the NRAS 5$^\prime$-UTR RNA G-quadruplex (PDB ID: 7SXP), shown in top and side views, together with predictions from OFoldNA and Boltz-1. \textbf{b,} Experimental structure of the \textit{Arabidopsis thaliana} ALOG domain--DNA complex (PDB ID: 8P5Q) and predictions from OFoldNA and AlphaFold 3. \textbf{c,} Experimental structure of the 53-nucleotide Lettuce C20G DNA aptamer bound to DFHO (PDB ID: 8FI1) and predictions from OFoldNA-SS2TS and Boltz-1. \textbf{d,e,} Protein--DNA (\textbf{d}) and protein-RNA (\textbf{e}) interfaces in the pre-strand-exchange IS621 recombinase complex containing bridge RNA, target DNA and donor DNA (PDB ID: 8WT6), together with predictions from OFoldNA-SS2TS and AlphaFold 3. OFoldNA-SS2TS denotes predictions generated using externally supplied SS information.}
    \label{fig:representative-structure-predictions}
\end{figure}

We next examined representative targets spanning challenging nucleic acid folds and protein--nucleic acid assemblies, both with and without external SS information. OFoldNA accurately recovered several challenging nucleic acid folds without external SS information. For the 22-nucleotide G8U variant of the G-quadruplex NRAS mRNA (PDB ID: 7SXP), most baselines failed to reproduce the compact stack of three G-tetrads, resulting in RMSDs of 9.8--13.3\,\AA{} and lDDT values of only 0.090--0.435. This failure extended to methods supplied with reference derived SS. RNAbpFlow, for example, still produced an RMSD above 10\,\AA{}, indicating that the model failed to effectively leverage the correct base-pairing information to recover the G-quadruplex fold. Boltz-1 reproduced the overall topology but showed substantial deviations in the sugar–phosphate backbone and loop regions, with a TM-score of 0.197 and an lDDT of 0.395. In contrast, OFoldNA achieved a TM-score of 0.394 and an lDDT of 0.750, recovered all non-Watson--Crick base pairs and reconstructed most of the stacking interactions that stabilize the G-quadruplex core (Fig.~\ref{fig:representative-structure-predictions}a). The same capability was evident for the 22-nucleotide parallel DNA G-quadruplex containing four G-tetrads (PDB ID: 8D78), for which OFoldNA reconstructed all four G-tetrads and all non-Watson--Crick base pairs, with a TM-score of 0.524 and an lDDT of 0.843.

The intrinsic prediction capability of OFoldNA also extended to protein--DNA assembly. In the ALOG domain--DNA complex of the \textit{Arabidopsis thaliana} transcription factor LSH3 (PDB ID: 8P5Q), a 129-residue ALOG domain binds a DNA duplex formed by two 17-nucleotide strands. OFoldNA recovered both the binding orientation of the duplex and its strand-specific protein contacts, achieving a mean DockQ of 0.517 and a mean iRMSD of 2.34\,\AA{} across the two protein--DNA interfaces (Fig.~\ref{fig:representative-structure-predictions}b). By contrast, the predictions from AF3 and Protenix-v1 both produced a mean DockQ of approximately 0.031 and a mean iRMSD of approximately 17.1\,\AA{}. Although their predicted DNA duplexes were close to the experimental structure in overall shape and spatial position, the duplex orientation and strand-specific contacts were mismatched, preventing recovery of the native protein--DNA interfaces. The advantage of OFoldNA therefore extended beyond approximating the overall complex geometry to resolving the correct interfacial assembly and native binding mode.

External SS information further enabled OFoldNA-SS2TS to recover difficult targets that were not recovered by any method under the standard prediction setting. The 53-nucleotide Lettuce C20G fluorogenic DNA aptamer (PDB ID: 8FI1), isolated by in vitro selection, adopts a compact tertiary structure formed by a four-way junction (4WJ), coaxial stacking and a central G-quadruplex containing two canonical G-tetrads. All predictions in the standard setting had TM-scores below 0.26 and RMSDs above 12\,\AA{}. When supplied with the reference base-pairing map, OFoldNA-SS2TS recovered the four-way-junction topology, the central G-quadruplex and the relative arrangement of the helical arms, yielding a TM-score of 0.481 and an RMSD of 2.975\,\AA{} (Fig.~\ref{fig:representative-structure-predictions}c). A more complex example is the pre-strand-exchange IS621 synaptic complex containing bridge RNA (bRNA), target DNA (tDNA) and donor DNA (dDNA; PDB ID: 8WT6). The assembly contains four IS621 protomers organized into two functional dimers, together with two bRNA modules that position the target and donor DNA substrates around the catalytic center. OFoldNA-SS2TS recovered all four protein--DNA interfaces, achieving a mean DockQ of 0.705, a mean iRMSD of 1.36\,\AA{} and a mean lRMSD of 3.83\,\AA{} (Fig.~\ref{fig:representative-structure-predictions}d). By comparison, AF3 and Protenix-v1 achieved mean DockQ scores of only 0.100 and 0.027. OFoldNA-SS2TS also recovered the assessed IS621--TBL protein--RNA interface, with a DockQ of 0.566, an iRMSD of 2.60\,\AA{} and an lRMSD of 4.41\,\AA{} (Fig.~\ref{fig:representative-structure-predictions}e); the corresponding DockQ scores from AF3 and Protenix-v1 were only 0.079 and 0.005. Thus, the supplied pairing information was translated beyond local base-pair recovery into the global topology and multicomponent organization of the complex. Importantly, key pairing constraints for this target can be inferred independently of the reference 3D structure: the bRNA stem–loop architecture was predicted from comparative sequence analysis, and candidate pairings with the target and donor DNA were identified through covariation analysis.

\subsection{OFoldNA learns generalizable RNA secondary structure representations}

To assess the generalizability of OFoldNA's learned SS representations, we evaluated OFoldNA-SS on the held-out in-distribution Test split and three out-of-distribution (OOD) splits of CHANRG~\cite{chen2026fair}. These OOD splits assess generalization to a held-out architectural regime (GenA), RNA clans absent from training (GenC), and families with limited reference-genome diversity (GenF). We compared OFoldNA-SS with three classes of RNA SS predictors: structured decoders (EternaFold~\cite{wayment2022eternafold} and RNAfold~\cite{lorenz2011viennarnafold}), direct neural predictors (BPfold~\cite{zhu2025bpfold} and SPOT-RNA~\cite{singh2019spot-rna}), and foundation-model predictors that use pretrained RNA language models (RiNALMo-Giga~\cite{penic2025rinalmo}, ERNIE-RNA~\cite{yin2025ernie} and structRFM~\cite{zhu2025fully}).

OFoldNA-SS achieved the highest mean OOD performance among the methods compared, with an $\mathrm{OOD}_{\mathrm{mean}}$ of 0.3898, representing an 8.0\% relative improvement over BPfold (0.3608; Table~\ref{tab:chanrg-ss-full-vs-50-200}). The corresponding $F_1$ scores were 0.3937 on GenA, 0.3723 on GenC and 0.4035 on GenF. On the in-distribution Test split, OFoldNA-SS achieved an $F_1$ of 0.4813, an 18.4\% relative improvement over BPfold (0.4065). By contrast, foundation-model predictors like RiNALMo achieved higher Test $F_1$ scores of 0.76 but substantially lower $\mathrm{OOD}_{\mathrm{mean}}$ scores of approximately 0.21, suggesting reduced generalization beyond the in-distribution Test split.

OFoldNA-SS retained the highest mean OOD performance after restricting all splits to sequences of 50--200 nucleotides, following the CHANRG protocol to reduce the influence of differences in sequence length composition. Under this restriction, OFoldNA-SS achieved an $\mathrm{OOD}_{\mathrm{mean}}$ of 0.4410, with base-pair $F_1$ scores of 0.5695 on GenA, 0.3504 on GenC and 0.4031 on GenF. Its Test $F_1$ reached 0.4899, remaining the highest among methods without pretrained RNA language models.

\input{tables/chanrg_ss.tex}
\subsection{Ablation analysis of secondary structure input and model components in OFoldNA}
We examined the contributions of SS input and model components to OFoldNA performance through ablation experiments on the FoldBench nucleic acid monomer test set.

\begin{figure}[!t]
    \centering
    \includegraphics[width=\linewidth]{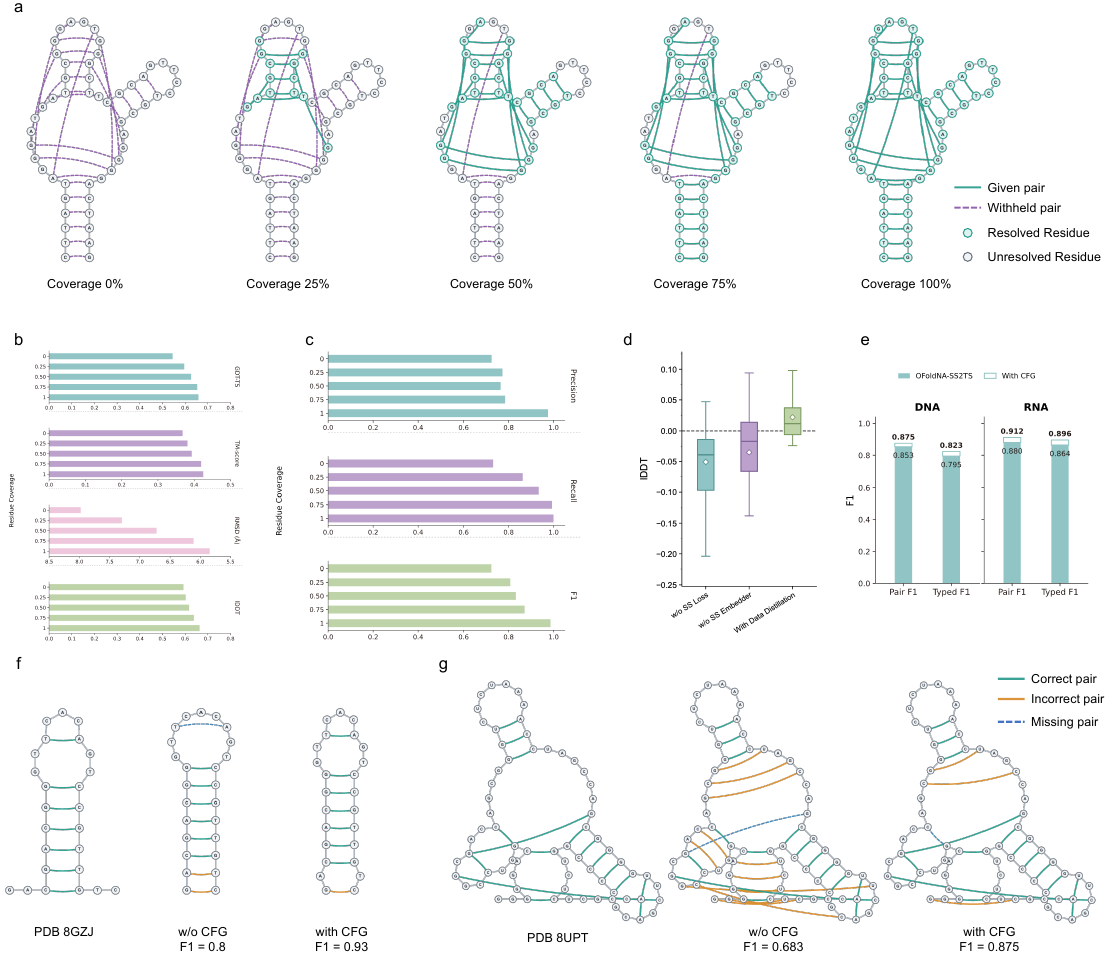}
    \caption{\textbf{Partial secondary structure information, core model ablations and optional extensions of OFoldNA.} \textbf{a,} Schematic of the resolved-residue coverage experiment. SS information is progressively revealed in locally connected regions of the pairing graph at nominal residue coverages of 0\%, 25\%, 50\%, 75\% and 100\%.  \textbf{b,} 3D structure accuracy at each resolved-residue coverage across the FoldBench nucleic acid monomer set ($n=29$ targets; 15 RNA and 14 DNA). \textbf{c,} Base-pair recovery at each coverage level, measured by precision, recall and $F_1$. \textbf{d,} Ablation analysis of OFoldNA without (w/o) the auxiliary SS loss or SS Embedder, together with the effect of data distillation, with performance measured by lDDT. \textbf{e,} Pair and Typed $F_1$ for OFoldNA-SS2TS predictions generated without or with CFG ($S=3$). Typed $F_1$ additionally requires matching FR3D interaction types. \textbf{f,g,} FR3D-derived SS for two representative targets with low $F_1$ scores without CFG: the DNA target 8GZJ (\textbf{f}) and the RNA target 8UPT (\textbf{g}).}
    \label{fig:partial-ss-distillation-cfg}
\end{figure}

To evaluate robustness to incomplete SS information, we progressively revealed locally connected regions of the pairing graph, corresponding to residue coverages of 0\%, 25\%, 50\%, 75\% and 100\% (Fig.~\ref{fig:partial-ss-distillation-cfg}a). OFoldNA-SS2TS showed a consistent gain in 3D structure accuracy as SS coverage increased. With pairing information available for only 25\% of residues, the mean GDT-TS increased from 0.545 to 0.596; at 75\% coverage, the mean GDT-TS and TM-score reached 0.653 and 0.419 (Fig.~\ref{fig:partial-ss-distillation-cfg}b). The SS prediction head showed a similar trend, with mean base-pair $F_1$ increasing from 0.724 without external SS input to 0.807 at 25\% coverage and 0.871 at 75\% coverage (Fig.~\ref{fig:partial-ss-distillation-cfg}c). These results show that the OFoldNA framework accommodates the varying availability of SS information encountered in practice, supporting prediction without external SS input and progressively benefiting from partial or complete pairing information.

The auxiliary SS loss and SS Embedder both contributed to model accuracy. Removing each component in separately trained variants reduced mean lDDT by 0.051 and 0.035. Distilled training data provided an additional gain, increasing mean lDDT by 0.022 relative to training without distillation (Fig.~\ref{fig:partial-ss-distillation-cfg}d). Classifier-free guidance (CFG) provides a complementary inference-time extension for OFoldNA-SS2TS. At a guidance strength of $S=3$, CFG increased Pair $F_1$ from 0.853 to 0.875 and Typed $F_1$ from 0.795 to 0.823 on DNA monomers, where Typed $F_1$ additionally requires the FR3D interaction type to match. On the RNA monomer targets, Pair $F_1$ increased from 0.880 to 0.912 and Typed $F_1$ from 0.864 to 0.896 (Fig.~\ref{fig:partial-ss-distillation-cfg}e). Representative examples illustrate that these gains arose from correction of both missed and spurious pairings. For 8GZJ, CFG increased both Pair and Typed $F_1$ from 0.800 to 0.933, recovering all seven reference pairs while reducing false-positive pairings (Fig.~\ref{fig:partial-ss-distillation-cfg}f). For 8UPT, both metrics increased from 0.683 to 0.875, accompanied by a reduction in false-positive pairs from 12 to 3 while retaining 14 of the 15 reference pairs (Fig.~\ref{fig:partial-ss-distillation-cfg}g).

%% file: tables/monomer_benchmark.tex
\begin{table}[t]
    \centering
    \caption{\textbf{Performance comparison on DNA and RNA monomer benchmarks.}}
    \label{tab:dna-rna-monomer-rank1}
    \resizebox{\linewidth}{!}{%
        \begin{tabular}{@{}cccccccccc@{}}
            \toprule
            \textbf{Dataset} & \textbf{Method} & \textbf{GDT-TS $\uparrow$} &
            \textbf{TM-score $\uparrow$} & \textbf{RMSD $\downarrow$} & \textbf{lDDT $\uparrow$} &
            \textbf{INF-all $\uparrow$} & \textbf{INF-stack $\uparrow$} & \textbf{INF-WC $\uparrow$} &
            \textbf{INF-nWC $\uparrow$} \\
            \midrule
            \multirow{9}{*}{DNA}
            & OFoldNA-SS2TS
            & \textbf{0.662} & \textbf{0.394} & \textbf{4.52}
            & \textbf{0.605} & \textbf{0.757} & \textbf{0.703}
            & \textbf{0.471} & \textbf{0.721} \\
            & 3dDNA
            & 0.302 & 0.152 & 12.93 & 0.258 & 0.431 & 0.462 & 0.399 & 0.240 \\
            \cmidrule(lr){2-10}
            & OFoldNA
            & \textbf{0.540} & \textbf{0.329} & \underline{7.65}
            & \textbf{0.560} & \textbf{0.688} & \textbf{0.694}
            & \underline{0.436} & \textbf{0.539} \\
            & Protenix-v1
            & 0.442 & 0.240 & 8.05 & \underline{0.464} & 0.614 & \underline{0.648} & 0.434 & 0.400 \\
            & AF3
            & \underline{0.492} & \underline{0.243} & \textbf{6.90}
            & \underline{0.464} & 0.608 & 0.624 & \underline{0.436} & 0.429 \\
            & Chai-1
            & 0.334 & 0.178 & 12.94 & 0.329 & 0.524 & 0.613 & 0.356 & 0.179 \\
            & Boltz-1
            & 0.447 & 0.239 & 8.87 & 0.307 & \underline{0.619} & 0.643
            & \textbf{0.452} & \underline{0.443} \\
            & RoseTTAFold2NA
            & 0.340 & 0.185 & 23.21 & 0.141 & 0.479 & 0.562 & 0.229 & 0.173 \\
            & DNAfold2
            & 0.268 & 0.130 & 24.26 & 0.009 & 0.341 & 0.375 & 0.289 & 0.000 \\
            \midrule
            \multirow{10}{*}{RNA}
            & OFoldNA-SS2TS
            & \textbf{0.657} & \textbf{0.454} & \textbf{7.08}
            & \textbf{0.719} & \textbf{0.861} & \textbf{0.847}
            & \textbf{0.838} & \textbf{0.737} \\
            & 3dRNA v2
            & 0.329 & 0.255 & 15.49 & 0.441 & 0.652 & 0.674 & 0.724 & 0.290 \\
            & RNAComposer
            & 0.361 & 0.268 & 12.36 & 0.497 & 0.702 & 0.703
            & \underline{0.825} & 0.366 \\
            & RNAbpFlow
            & \underline{0.507} & \underline{0.345} & \underline{9.06} & \underline{0.636}
            & \underline{0.727} & \underline{0.736} & 0.817 & \underline{0.436} \\
            \cmidrule(lr){2-10}
            & OFoldNA
            & \textbf{0.550} & \textbf{0.404} & \textbf{8.28}
            & \textbf{0.623} & \textbf{0.785} & \textbf{0.802}
            & 0.779 & \underline{0.537} \\
            & Protenix-v1
            & \underline{0.541} & \underline{0.392} & \underline{9.00} & \underline{0.614} & 0.775 & \underline{0.793}
            & \textbf{0.813} & \textbf{0.576} \\
            & AF3
            & 0.467 & 0.354 & 11.36 & 0.582 & 0.749 & 0.783 & \underline{0.787} & 0.469 \\
            & Boltz-1
            & 0.498 & 0.351 & 11.45 & 0.464 & \underline{0.778} & \underline{0.793} & 0.786 & 0.490 \\
            & Chai-1
            & 0.385 & 0.294 & 14.93 & 0.457 & 0.669 & 0.729 & 0.587 & 0.350 \\
            & RoseTTAFold2NA
            & 0.345 & 0.269 & 17.45 & 0.197 & 0.637 & 0.686 & 0.579 & 0.325 \\
            \bottomrule
        \end{tabular}%
    }\par
    \vspace{2pt}
    \parbox{\linewidth}{\footnotesize\raggedright
        Within each dataset, methods above the internal rule use secondary structure inputs, whereas methods below the rule do not. Best and second-best results are shown in bold and underlined, respectively, and are determined separately for the two input settings.\par}
\end{table}

%% file: tables/interface_full.tex
\begin{table}[t]
    \centering
    \caption{\textbf{Performance on the full FoldBench protein--DNA and protein--RNA interface subsets.}}
    \label{tab:protein-na-interface-rank1-all-lengths}
    \begin{tabular*}{\linewidth}{@{\extracolsep{\fill}}clrrrr@{}}
        \toprule
        \textbf{Dataset} & \textbf{Method} & \textbf{DockQ $\uparrow$} &
        \textbf{iRMSD $\downarrow$} & \textbf{lRMSD $\downarrow$} & \textbf{lDDT $\uparrow$} \\
        \midrule
        \multirow{7}{*}{\textbf{Protein--DNA}}
        & OFoldNA-SS2TS
        & \textbf{0.617} & \textbf{3.707} & \textbf{8.322} & \textbf{0.856} \\
        & OFoldNA
        & \underline{0.604} & \underline{4.117} & \underline{9.954} & \underline{0.851} \\
        & AF3
        & 0.596 & 4.514 & 10.432 & 0.839 \\
        & Protenix-v1
        & 0.584 & 4.625 & 10.557 & 0.847 \\
        & Boltz-1
        & 0.541 & 5.048 & 12.230 & 0.805 \\
        & Chai-1
        & 0.541 & 4.611 & 11.015 & 0.811 \\
        & RF2NA
        & 0.286 & 13.834 & 26.312 & 0.587 \\
        \midrule
        \multirow{7}{*}{\textbf{Protein--RNA}}
        & OFoldNA-SS2TS
        & \textbf{0.420} & \textbf{6.100} & \textbf{14.238} & \textbf{0.774} \\
        & OFoldNA
        & \underline{0.400} & 7.294 & \underline{16.451} & 0.769 \\
        & AF3
        & 0.360 & \underline{6.807} & 17.915 & 0.771 \\
        & Protenix-v1
        & 0.395 & 7.013 & 17.993 & \underline{0.772} \\
        & Boltz-1
        & 0.296 & 8.592 & 20.798 & 0.679 \\
        & Chai-1
        & 0.239 & 9.789 & 27.346 & 0.719 \\
        & RF2NA
        & 0.096 & 36.259 & 56.202 & 0.427 \\
        \bottomrule
    \end{tabular*}\par
    \vspace{2pt}
    \parbox{\linewidth}{\footnotesize\raggedright
        Best and second-best results within each dataset are shown in bold and underlined.\par}
\end{table}

%% file: tables/chanrg_ss.tex
\begin{table}[t]
    \centering
    \caption{\textbf{Comparison of base-pair $F_1$ on the full CHANRG evaluation splits and their 50--200-nucleotide length-controlled subsets.}}
    \label{tab:chanrg-ss-full-vs-50-200}
    \resizebox{\linewidth}{!}{%
        \begin{tabular}{@{}lcrrrrrrrrrr@{}}
            \toprule
            \multirow{2}{*}{\textbf{Method}} &
            \multirow{2}{*}{\textbf{Type}} &
            \multicolumn{5}{c}{\textbf{Full splits}} &
            \multicolumn{5}{c}{\textbf{50--200 nt (length-controlled)}} \\
            \cmidrule(lr){3-7}\cmidrule(lr){8-12}
            & & \textbf{Test $\uparrow$} & \textbf{GenA $\uparrow$} & \textbf{GenC $\uparrow$} & \textbf{GenF $\uparrow$} & $\mathbf{OOD}_{\mathbf{mean}}\,\uparrow$
            & \textbf{Test $\uparrow$} & \textbf{GenA $\uparrow$} & \textbf{GenC $\uparrow$} & \textbf{GenF $\uparrow$} & $\mathbf{OOD}_{\mathbf{mean}}\,\uparrow$ \\
            \midrule
            OFoldNA-SS & DL
            & 0.4813 & \textbf{0.3937} & \underline{0.3723}
            & \textbf{0.4035} & \textbf{0.3898}
            & 0.4899 & \textbf{0.5695} & \underline{0.3504}
            & \textbf{0.4031} & \textbf{0.4410} \\
            BPfold & DL
            & 0.4065 & \underline{0.3876} & \textbf{0.4137}
            & 0.2811 & \underline{0.3608}
            & 0.4463 & \underline{0.5687} & \textbf{0.3956}
            & 0.2773 & \underline{0.4139} \\
            SPOT-RNA & DL
            & 0.3761 & 0.3477 & 0.3378 & 0.2524 & 0.3126
            & 0.4123 & 0.5304 & 0.3209 & 0.2481 & 0.3664 \\
            EternaFold & SD
            & 0.3189 & 0.3171 & 0.3108 & \underline{0.2913} & 0.3064
            & 0.3345 & 0.4608 & 0.2882 & \underline{0.2874} & 0.3455 \\
            RNAfold & SD
            & 0.3013 & 0.2802 & 0.3011 & 0.2846 & 0.2887
            & 0.3203 & 0.4193 & 0.2799 & 0.2812 & 0.3268 \\
            RiNALMo-Giga & FM
            & \textbf{0.7579} & 0.2509 & 0.1651 & 0.2260 & 0.2140
            & \textbf{0.7832} & 0.3565 & 0.1525 & 0.2298 & 0.2463 \\
            ERNIE-RNA & FM
            & \underline{0.7061} & 0.3308 & 0.2803 & 0.2309 & 0.2807
            & \underline{0.7300} & 0.5155 & 0.2699 & 0.2340 & 0.3398 \\
            structRFM & FM
            & 0.4236 & 0.3781 & 0.3033 & 0.2557 & 0.3124
            & 0.4619 & 0.5514 & 0.2873 & 0.2526 & 0.3638 \\
            \bottomrule
        \end{tabular}%
    }\par
    \vspace{2pt}
    \parbox{\linewidth}{\footnotesize\raggedright
        $\mathrm{OOD}_{\mathrm{mean}}$ is the unweighted mean across GenA, GenC and GenF. Best and second-best results are shown in bold and underlined. DL, direct neural predictor; SD, structured decoder; FM, foundation-model predictor.\par}
\end{table}

%% file: sections/05_discussion.tex
\section{Discussion}

In this work, we developed OFoldNA, a unified all-atom diffusion framework for nucleic acid monomer folding and protein--nucleic acid co-folding. OFoldNA treats base-pairing as both an internally learned representation and an optional external condition rather than as a prerequisite for 3D prediction. Benchmarking on FoldBench showed that OFoldNA achieved leading performance in both nucleic acid monomer folding and protein--nucleic acid co-folding, with particularly strong improvements on DNA monomers and challenging interfaces involving longer DNA chains. The G-quadruplex and ALOG domain--DNA examples further show that these gains extend beyond global fold similarity to the recovery of noncanonical interaction networks and strand-specific protein--nucleic acid binding modes.

When supplied with pairing maps extracted from the reference structures, OFoldNA-SS2TS further improved prediction accuracy. Predictions for the Lettuce aptamer and the IS621 synaptic complex demonstrate that OFoldNA-SS2TS translates supplied pairing relationships into tertiary folds and multicomponent assemblies rather than merely reproducing local base pairs. As a dedicated SS predictor, OFoldNA-SS achieved the highest mean out-of-distribution accuracy on CHANRG, which supports the generalizability of its learned SS representations.  

However, predictions involving longer nucleic acid chains remain challenging, owing in part to the limited number of experimentally determined structures of nucleic acids and their complexes available for training. The current framework also does not explicitly model how ion concentrations and other solution conditions influence nucleic acid folding and conformational equilibria. Incorporating these environmental variables may enable more accurate prediction of condition-dependent structures and conformational states. OFoldNA-SS2TS supports different levels of SS availability, but external SS information must still be provided as pairing maps. Future work could extend conditioning beyond base-pairing information by directly incorporating nucleotide-level reactivity profiles from SHAPE or DMS probing, allowing a broader range of experimental signals to guide all-atom structure prediction, as well as non-canonical base pairing~\cite{NCfold} for more accurate secondary structure geometry. Despite these limitations, OFoldNA provides a unified framework for predicting base pairing internally and using partial or complete external pairing maps to guide all-atom folding and co-folding.

%% file: sections/03_methods.tex
\section{Methods}

\subsection{OFoldNA architecture}

OFoldNA builds on the AlphaFold 3 architecture implemented in Protenix. Biomolecular inputs are encoded as single and pair representations, which are iteratively updated by the Pairformer trunk through recycling and passed to a diffusion module for all-atom structure generation. We introduce an SS embedder that incorporates pairing-map features into the pair representation and an auxiliary head that predicts base-pairing probabilities from the final pair representation. The same architecture also supports standalone SS prediction, a mode termed OFoldNA-SS. OFoldNA predicts 3D structures without an external pairing map, whereas OFoldNA-SS2TS conditions structure generation on a complete or partial pairing map and optionally applies SS-conditioned classifier-free guidance during diffusion sampling.

\subsection{SS conditioning module}
Reference pairing maps for training were extracted from experimentally determined structures using FR3D~\cite{sarver2008fr3d}. For an input of $N$ tokens, let $\mathbf{z}\in\mathbb{R}^{N\times N\times c_z}$ denote the trunk pair representation. The $k$th SS condition is represented by a symmetric matrix $\mathbf{S}^{(k)}\in\mathbb{R}^{N\times N}$ and a token-level validity mask $\mathbf{m}^{(k)}\in\{0,1\}^{N}$. Its discrete entries are defined as
\begin{equation}
S_{ij}^{(k)}=
\begin{cases}
1, & i\text{ and }j\text{ are a known pair},\\
0, & i\text{ and }j\text{ are a known non-pair},\\
-1, & \text{their pairing state is unknown}.
\end{cases}
\end{equation}
The pairwise validity mask is the outer product of the token-level mask:
\begin{equation}
M_{ij}^{(k)}=m_i^{(k)}m_j^{(k)}.
\end{equation}

Each token type is represented by a one-hot vector $\mathbf{a}_i$ over the model's unified residue vocabulary. For token pair $(i,j)$, the masked SS input feature is
\begin{equation}
\mathbf{h}_{ij}^{(k)}
=M_{ij}^{(k)}
\left[
\mathbf{a}_i,\,
\mathbf{a}_j,\,
S_{ij}^{(k)},\,
M_{ij}^{(k)}
\right].
\end{equation}

Here, the brackets denote feature concatenation. Pairs outside the valid region ($M_{ij}^{(k)}=0$) are zeroed, whereas valid pairs with $S_{ij}^{(k)}=-1$ retain an explicit UNKNOWN state. This distinction allows the model to separate missing SS coverage from an unknown pairing state within the covered region.

During training, we randomly mask the conditioning matrix to improve robustness to incomplete SS input. For a subset of examples, all pairing states are set to UNKNOWN:
\begin{equation}
\widetilde{S}_{ij}^{(k)}=-1,\qquad \forall i,j.
\end{equation}

Otherwise, a subset of known pairs and non-pairs is set to $-1$. Mask positions are sampled from the strict upper triangle and mirrored to the lower triangle so that $(i,j)$ and $(j,i)$ are always masked together. Only the conditioning map is masked; the auxiliary prediction head is supervised against the unmasked SS labels.

After layer normalization, the pair representation is linearly projected and added to a linear projection of the SS input:
\begin{equation}
\mathbf{z}_{\mathrm{SS,in}}^{(k)}
=W_z\operatorname{LN}(\mathbf{z})+W_h\mathbf{h}^{(k)}.
\end{equation}

Here, $\operatorname{LN}$ denotes layer normalization. A lightweight Pairformer stack updates only the pair representation. Layer normalization, ReLU activation and a final linear projection then produce a residual with the same channel dimension as the trunk pair representation:
\begin{equation}
\Delta\mathbf{z}^{(k)}
=W_o\operatorname{ReLU}
\left(
\operatorname{LN}
\left[
\operatorname{Pairformer}
\left(\mathbf{z}_{\mathrm{SS,in}}^{(k)}\right)
\right]
\right).
\end{equation}

For $K$ input SS maps, each condition is encoded independently and the resulting residuals are averaged:
\begin{equation}
\Delta\mathbf{z}
=\frac{1}{K}\sum_{k=1}^{K}\Delta\mathbf{z}^{(k)}.
\end{equation}

At each recycling cycle, the fused SS residual is added to the current pair representation:
\begin{equation}
\mathbf{z}\leftarrow\mathbf{z}+\Delta\mathbf{z}.
\end{equation}
The updated $\mathbf{z}$ is then passed to the main Pairformer trunk, allowing the SS condition to influence every recycling cycle.

\subsection{SS prediction and auxiliary supervision}

An SS prediction head decodes base-pairing information from the final pair representation $\mathbf{z}$. The head first applies layer normalization followed by a linear projection:
\begin{equation}
\mathbf{z}'=W_p\operatorname{LN}(\mathbf{z}).
\end{equation}
The projected pair features are then symmetrized across the two token axes:
\begin{equation}
\overline{\mathbf{z}}_{ij}
=\frac{1}{2}\left(\mathbf{z}'_{ij}+\mathbf{z}'_{ji}\right).
\end{equation}

A feed-forward network maps each symmetric pair feature to a logit $\ell_{ij}$. Because symmetrization precedes the feed-forward network, $\ell_{ij}=\ell_{ji}$, and the corresponding pairing probability is $p_{ij}=\sigma(\ell_{ij})$.

We train the head using masked binary cross-entropy. Let $Y_{ij}\in\{-1,0,1\}$ denote an SS label and $\mathbf{m}$ its token-level validity mask. The pairwise validity mask and known-state mask are
\begin{equation}
M_{ij}=m_i m_j,\qquad
U_{ij}=\mathbb{1}\!\left[Y_{ij}\ne-1\right].
\end{equation}

Their product, $W_{ij}=M_{ij}U_{ij}$, defines the combined loss mask, and the SS loss is
\begin{equation}
\mathcal{L}_{\mathrm{SS}}
=
-\frac{1}{\epsilon+\sum_{ij}W_{ij}}
\sum_{ij}W_{ij}
\left[
Y_{ij}\log p_{ij}
+(1-Y_{ij})\log(1-p_{ij})
\right].
\end{equation}

Entries labeled UNKNOWN are excluded by $U$ and therefore do not contribute as non-pairs. The total training objective is
\begin{equation}
\mathcal{L}
=\mathcal{L}_{\mathrm{base}}
+\lambda_{\mathrm{SS}}\mathcal{L}_{\mathrm{SS}},
\end{equation}
where $\epsilon$ is a small constant for numerical stability and $\lambda_{\mathrm{SS}}$ weights the auxiliary SS loss. During training, the SS head is applied to the pair representation from the final recycling cycle and optimized jointly with the 3D structure-prediction objectives.

\subsection{SS classifier-free guidance}

During sampling, classifier-free guidance (CFG) is optional to control the influence of the SS condition on coordinate diffusion. The conditional branch uses the input SS matrix $\mathbf{S}_{\mathrm{cond}}$, whereas the all-UNKNOWN reference branch preserves the same token-validity mask but sets all pairing states to UNKNOWN, that is, $\mathbf{S}_{\mathrm{ref}}=-\mathbf{1}$. The reference branch therefore represents the absence of pairing-state information rather than a map in which all token pairs are non-pairs. The trunk is run separately for the two branches, and the resulting single and pair representations are processed by the shared diffusion denoiser. At each diffusion step, the two branches receive the same noisy coordinates $\mathbf{x}_t$ and noise level $t$, producing $\widehat{\mathbf{x}}_{\mathrm{cond}}$ and $\widehat{\mathbf{x}}_{\mathrm{unknown}}$. These predictions are combined as
\begin{equation}
\widehat{\mathbf{x}}_{\mathrm{CFG}}
=
\widehat{\mathbf{x}}_{\mathrm{unknown}}
+S\left(
\widehat{\mathbf{x}}_{\mathrm{cond}}
-\widehat{\mathbf{x}}_{\mathrm{unknown}}
\right),
\end{equation}
where $S$ is the scalar guidance scale. When $S>1$, the difference between the conditional and all-UNKNOWN branches is amplified, increasing the influence of the SS condition on the diffusion trajectory. Guidance is applied directly to the denoised coordinate predictions and requires no separately trained classifier.

\subsection{Training data}
Experimental structures released in the Protein Data Bank (PDB)~\cite{berman2000pdb} before 30 September 2021 were collected for training. Entries were retained if they had a resolution of $9.0\,\text{\AA}$ or better and contained at least one DNA or RNA polymer chain. The resulting data index contained 11,758 PDB mmCIF files.

Following Boltz-2~\cite{passaro2025boltz2}, we constructed the protein--DNA distillation data by sampling ten single-stranded DNA motifs per protein sequence using JASPAR 2024 CORE profiles matched to two high-throughput SELEX datasets. Each motif and its reverse complement were supplied to Boltz-1 together with the corresponding protein sequence. Predictions were retained if the predicted distance error (PDE) was no greater than 2.0 and the maximum interface predicted distance error (iPDE) was no greater than 1.0. The minimum interface predicted TM-score (ipTM) was required to be at least 0.70.

For protein--RNA self-distillation, candidate pairs from CISBP-RNA and an additional set of sequence-defined protein--RNA pairs were supplied to OFoldNA to generate pseudo-labeled complex structures. Predictions with model-reported clashes were excluded. Among the remaining predictions, ranking score and ipTM values were both required to be at least 0.70. The mean interface predicted aligned error (PAE) was required to be no greater than $5.0\,\text{\AA}$, with the 90th percentile of the interface PAE values no greater than $8.0\,\text{\AA}$. Each prediction was also required to contain at least three protein--RNA residue contacts.

For RNA monomer self-distillation, CHANRG sequences and their corresponding SS were supplied to OFoldNA-SS2TS. Predictions were retained if pLDDT exceeded 0.60 and the base-pair $F_1$ score between the supplied and structure-derived SS exceeded 0.70. For the distillation-augmented variant, predictions retained from all three molecular classes were added to the experimental structures as training examples.

\subsection{Model training}

All model parameters were jointly optimized for 40,000 steps using Adam with a learning rate of $10^{-3}$, $\beta_1=0.9$, $\beta_2=0.95$ and a weight decay of $10^{-8}$. We followed the AlphaFold 3 learning-rate schedule, using 1,000 warm-up steps and clipping the gradient norm at 10. An exponential moving average of the model parameters was maintained with a decay rate of 0.999. Training on 16 NVIDIA A800 GPUs took approximately 6 days.

\subsection{Inference and baseline methods}

Except for RNAComposer, 3dRNA v2 and 3dDNA, all baseline methods were obtained from their official code repositories and run locally using the inference procedures recommended by the respective authors. RNAComposer, 3dRNA v2 and 3dDNA were accessed through their official web servers, with 10 candidate structures generated per target for each method. DNAfold2 was run independently 25 times per target. All other locally run methods used the 5 FoldBench random seeds (42, 66, 101, 2024 and 8888) and generated 5 predictions per seed, yielding 25 candidate structures per target. The inference entry point of RoseTTAFold2NA was adapted to implement the same multi-seed sampling protocol. OFoldNA was run with 10 recycling iterations and 200 diffusion steps.

Reference-derived base-pairing information was provided as an additional input to OFoldNA-SS2TS, RNAbpFlow, RNAComposer, 3dRNA v2 and 3dDNA. For the FoldBench evaluation of RNAbpFlow, lDDT was computed with stereochemical checks disabled (\texttt{lddt\_no\_stereochecks: true}). Locally generated predictions were ranked using the confidence or ranking score provided by each method. For web-server methods, we used the server-provided ordering.

\subsection{Evaluation metrics}

Structure predictions were evaluated using the FoldBench evaluation pipeline. For nucleic acid monomers, global structural accuracy was quantified using GDT-TS, TM-score and root-mean-square deviation (RMSD), whereas local accuracy was assessed using lDDT. Interaction network fidelity (INF) was used to assess the recovery of base-pairing and stacking interactions, including INF-all, INF-stack, INF-WC and INF-nWC. Predictions with a TM-score greater than 0.45 were considered to have recovered the correct global fold.

Protein--nucleic acid interfaces were evaluated using DockQ, interface RMSD (iRMSD), ligand RMSD (lRMSD) and lDDT. DockQ quality was categorized as acceptable ($0.23 \leq \mathrm{DockQ}<0.49$), medium ($0.49 \leq \mathrm{DockQ}<0.80$) or high ($\mathrm{DockQ}\geq0.80$). SS predictions were evaluated using base-pair precision, recall and $F_1$. Pair $F_1$ required the paired residue indices to match the reference, whereas Typed $F_1$ additionally required the FR3D interaction type to match.

%% file: appendix/01_appendix.tex
\section{Supplementary Information}

\input{tables/predicted_ss_monomer.tex}
\input{tables/interface_ge20nt.tex}

%% file: tables/predicted_ss_monomer.tex
\begin{table}[H]
    \centering
    \caption{\textbf{Performance on FoldBench RNA monomers using internally predicted and externally predicted secondary-structure maps.}}
    \label{tab:rna-monomer-predicted-ss-rank1}
    \resizebox{\linewidth}{!}{%
        \begin{tabular}{@{}lccccccccc@{}}
            \toprule
            \textbf{SS setting} & \textbf{GDT-TS $\uparrow$} &
            \textbf{TM-score $\uparrow$} & \textbf{RMSD $\downarrow$} &
            \textbf{lDDT $\uparrow$} & \textbf{INF-all $\uparrow$} &
            \textbf{INF-stack $\uparrow$} & \textbf{INF-WC $\uparrow$} &
            \textbf{INF-nWC $\uparrow$} & \textbf{ base-pair $F_1$ $\uparrow$} \\
            \midrule
            OFoldNA-SS
            & \textbf{0.550} & \textbf{0.404} & \textbf{8.28}
            & \textbf{0.623} & \textbf{0.785} & \textbf{0.802}
            & 0.779 & \textbf{0.537} & \textbf{0.368} \\
            SPOT-RNA
            & \underline{0.521} & \underline{0.389} & 10.57
            & \underline{0.601} & \underline{0.753} & \underline{0.779}
            & \textbf{0.807} & \underline{0.489} & \underline{0.347} \\
            EternaFold
            & 0.514 & 0.367 & \underline{10.14}
            & 0.570 & 0.729 & 0.757
            & \underline{0.782} & 0.443 & 0.320 \\
            RNAfold
            & 0.489 & 0.367 & 10.62
            & 0.574 & 0.717 & 0.754
            & 0.729 & 0.483 & 0.265 \\
            BPfold
            & 0.509 & 0.357 & 11.25
            & 0.573 & 0.739 & 0.777
            & 0.753 & 0.474 & 0.305 \\
            RiNALMo
            & 0.479 & 0.363 & 11.60
            & 0.543 & 0.703 & 0.747
            & 0.717 & 0.396 & 0.267 \\
            \bottomrule
        \end{tabular}%
    }\par
    \vspace{2pt}
    \parbox{\linewidth}{\footnotesize\raggedright
        Values are mean rank-1 results. Best and second-best results across the six settings are shown in bold and underlined, respectively. \par}
\end{table}

%% file: tables/interface_ge20nt.tex
\begin{table}[H]
    \centering
    \caption{\textbf{Performance comparison on FoldBench protein--DNA and protein--RNA interfaces containing nucleic acid chains of at least 20 nucleotides.}}
    \label{tab:protein-na-interface-rank1}
    \begin{tabular*}{\linewidth}{@{\extracolsep{\fill}}clrrrr@{}}
        \toprule
        \textbf{Dataset} & \textbf{Method} & \textbf{DockQ $\uparrow$} &
        \textbf{iRMSD $\downarrow$} & \textbf{lRMSD $\downarrow$} & \textbf{lDDT $\uparrow$} \\
        \midrule
        \multirow{7}{*}{\textbf{Protein--DNA}}
        & OFoldNA-SS2TS
        & \textbf{0.351} & \textbf{8.805} & \textbf{18.934} & \textbf{0.788} \\
        & OFoldNA
        & \underline{0.286} & \underline{10.080} & \underline{25.971} & 0.764 \\
        & AlphaFold3
        & 0.238 & 12.611 & 28.619 & \underline{0.770} \\
        & Protenix-v1
        & 0.215 & 12.198 & 28.167 & 0.770 \\
        & Boltz-1
        & 0.147 & 13.740 & 35.037 & 0.687 \\
        & Chai-1
        & 0.203 & 10.272 & 26.220 & 0.713 \\
        & RoseTTAFold2NA
        & 0.069 & 18.103 & 44.515 & 0.422 \\
        \midrule
        \multirow{7}{*}{\textbf{Protein--RNA}}
        & OFoldNA-SS2TS
        & \textbf{0.408} & \textbf{7.761} & \textbf{17.582} & \textbf{0.715} \\
        & OFoldNA
        & \underline{0.362} & 10.146 & \underline{22.116} & 0.700 \\
        & AlphaFold3
        & 0.287 & \underline{8.997} & 24.736 & 0.704 \\
        & Protenix-v1
        & 0.359 & 9.433 & 25.386 & \underline{0.711} \\
        & Boltz-1
        & 0.274 & 12.045 & 29.196 & 0.569 \\
        & Chai-1
        & 0.145 & 14.035 & 41.402 & 0.636 \\
        & RoseTTAFold2NA
        & 0.036 & 42.961 & 66.259 & 0.281 \\
        \bottomrule
    \end{tabular*}\par
    \vspace{2pt}
    \parbox{\linewidth}{\footnotesize\raggedright
        Values are means over each method's available rank-1 predictions from the filtered sets of 58 protein--DNA and 33 protein--RNA interfaces. Best and second-best results within each dataset are shown in bold and underlined, respectively.\par}
\end{table}

%% file: references.bib
@article{crick1970central,
  title={Central dogma of molecular biology},
  author={Crick, Francis},
  journal={Nature},
  volume={227},
  number={5258},
  pages={561--563},
  year={1970},
  publisher={Nature Publishing Group UK London}
}

@article{wang2023dynamic,
  title={Dynamic alternative DNA structures in biology and disease},
  author={Wang, Guliang and Vasquez, Karen M},
  journal={Nature Reviews Genetics},
  volume={24},
  number={4},
  pages={211--234},
  year={2023},
  publisher={Nature Publishing Group UK London}
}

@article{wulfridge2024intertwining,
  title={Intertwining roles of R-loops and G-quadruplexes in DNA repair, transcription and genome organization},
  author={Wulfridge, Phillip and Sarma, Kavitha},
  journal={Nature cell biology},
  volume={26},
  number={7},
  pages={1025--1036},
  year={2024},
  publisher={Nature Publishing Group UK London}
}

@article{petermann2022sources,
  title={Sources, resolution and physiological relevance of R-loops and RNA--DNA hybrids},
  author={Petermann, Eva and Lan, Li and Zou, Lee},
  journal={Nature reviews Molecular cell biology},
  volume={23},
  number={8},
  pages={521--540},
  year={2022},
  publisher={Nature Publishing Group UK London}
}

@article{cao2024identification,
  title={Identification of RNA structures and their roles in RNA functions},
  author={Cao, Xinang and Zhang, Yueying and Ding, Yiliang and Wan, Yue},
  journal={Nature Reviews Molecular Cell Biology},
  volume={25},
  number={10},
  pages={784--801},
  year={2024},
  publisher={Nature Publishing Group UK London}
}

@article{ganser2019roles,
  title={The roles of structural dynamics in the cellular functions of RNAs},
  author={Ganser, Laura R and Kelly, Megan L and Herschlag, Daniel and Al-Hashimi, Hashim M},
  journal={Nature reviews Molecular cell biology},
  volume={20},
  number={8},
  pages={474--489},
  year={2019},
  publisher={Nature Publishing Group UK London}
}

@article{dykstra2022engineering,
  title={Engineering synthetic RNA devices for cell control},
  author={Dykstra, Peter B and Kaplan, Matias and Smolke, Christina D},
  journal={Nature Reviews Genetics},
  volume={23},
  number={4},
  pages={215--228},
  year={2022},
  publisher={Nature Publishing Group UK London}
}

@article{spitale2023probing,
  title={Probing the dynamic RNA structurome and its functions},
  author={Spitale, Robert C and Incarnato, Danny},
  journal={Nature Reviews Genetics},
  volume={24},
  number={3},
  pages={178--196},
  year={2023},
  publisher={Nature Publishing Group UK London}
}

@article{ducoli2026bridging,
  title={Bridging technical innovation and computational advances in studies of RNA--protein assemblies},
  author={Ducoli, Luca and Srinivasan, Suhas and Amjadi, Eimon and Khavari, Paul A},
  journal={Nature Reviews Genetics},
  volume={27},
  number={6},
  pages={462--484},
  year={2026},
  publisher={Nature Publishing Group UK London}
}

@article{childs2022targeting,
  title={Targeting RNA structures with small molecules},
  author={Childs-Disney, Jessica L and Yang, Xueyi and Gibaut, Quentin MR and Tong, Yuquan and Batey, Robert T and Disney, Matthew D},
  journal={Nature Reviews Drug Discovery},
  volume={21},
  number={10},
  pages={736--762},
  year={2022},
  publisher={Nature Publishing Group UK London}
}

@article{zhang2022advances,
  title={Advances and opportunities in RNA structure experimental determination and computational modeling},
  author={Zhang, Jinsong and Fei, Yuhan and Sun, Lei and Zhang, Qiangfeng Cliff},
  journal={Nature methods},
  volume={19},
  number={10},
  pages={1193--1207},
  year={2022},
  publisher={Nature Publishing Group US New York}
}

@article{wang2026integrated,
  title={Integrated experimental and AI innovations for RNA structure determination},
  author={Wang, Wenkai and Su, Baoquan and Peng, Zhenling and Yang, Jianyi},
  journal={Nature Biotechnology},
  volume={44},
  number={2},
  pages={205--214},
  year={2026},
  publisher={Nature Publishing Group US New York}
}

@article{berman2000pdb,
  title={The protein data bank},
  author={Berman, Helen M and Westbrook, John and Feng, Zukang and Gilliland, Gary and Bhat, Talapady N and Weissig, Helge and Shindyalov, Ilya N and Bourne, Philip E},
  journal={Nucleic acids research},
  volume={28},
  number={1},
  pages={235--242},
  year={2000},
  publisher={Oxford University Press}
}

@article{siegfried2014rna,
  title={RNA motif discovery by SHAPE and mutational profiling (SHAPE-MaP)},
  author={Siegfried, Nathan A and Busan, Steven and Rice, Greggory M and Nelson, Julie AE and Weeks, Kevin M},
  journal={Nature methods},
  volume={11},
  number={9},
  pages={959--965},
  year={2014},
  publisher={Nature Publishing Group US New York}
}

@article{spitale2015structural,
  title={Structural imprints in vivo decode RNA regulatory mechanisms},
  author={Spitale, Robert C and Flynn, Ryan A and Zhang, Qiangfeng Cliff and Crisalli, Pete and Lee, Byron and Jung, Jong-Wha and Kuchelmeister, Hannes Y and Batista, Pedro J and Torre, Eduardo A and Kool, Eric T and others},
  journal={Nature},
  volume={519},
  number={7544},
  pages={486--490},
  year={2015},
  publisher={Nature Publishing Group UK London}
}

@article{zubradt2017dms,
  title={DMS-MaPseq for genome-wide or targeted RNA structure probing in vivo},
  author={Zubradt, Meghan and Gupta, Paromita and Persad, Sitara and Lambowitz, Alan M and Weissman, Jonathan S and Rouskin, Silvi},
  journal={Nature methods},
  volume={14},
  number={1},
  pages={75--82},
  year={2017},
  publisher={Nature Publishing Group US New York}
}

@article{chambers2015high,
  title={High-throughput sequencing of DNA G-quadruplex structures in the human genome},
  author={Chambers, Vicki S and Marsico, Giovanni and Boutell, Jonathan M and Di Antonio, Marco and Smith, Geoffrey P and Balasubramanian, Shankar},
  journal={Nature biotechnology},
  volume={33},
  number={8},
  pages={877--881},
  year={2015},
  publisher={Nature Publishing Group US New York}
}

@article{zeraati2018motif,
  title={I-motif DNA structures are formed in the nuclei of human cells},
  author={Zeraati, Mahdi and Langley, David B and Schofield, Peter and Moye, Aaron L and Rouet, Romain and Hughes, William E and Bryan, Tracy M and Dinger, Marcel E and Christ, Daniel},
  journal={Nature chemistry},
  volume={10},
  number={6},
  pages={631--637},
  year={2018},
  publisher={Nature Publishing Group UK London}
}

@article{kouzine2017permanganate,
  title={Permanganate/S1 nuclease footprinting reveals non-B DNA structures with regulatory potential across a mammalian genome},
  author={Kouzine, Fedor and Wojtowicz, Damian and Baranello, Laura and Yamane, Arito and Nelson, Steevenson and Resch, Wolfgang and Kieffer-Kwon, Kyong-Rim and Benham, Craig J and Casellas, Rafael and Przytycka, Teresa M and others},
  journal={Cell systems},
  volume={4},
  number={3},
  pages={344--356},
  year={2017},
  publisher={Elsevier}
}

@article{bedrat2016re,
  title={Re-evaluation of G-quadruplex propensity with G4Hunter},
  author={Bedrat, Amina and Lacroix, Laurent and Mergny, Jean-Louis},
  journal={Nucleic acids research},
  volume={44},
  number={4},
  pages={1746--1759},
  year={2016},
  publisher={Oxford University Press}
}

@article{yu2024seeker,
  title={iM-Seeker: a webserver for DNA i-motifs prediction and scoring via automated machine learning},
  author={Yu, Haopeng and Li, Fan and Yang, Bibo and Qi, Yiman and Guneri, Dilek and Chen, Wenqian and Waller, Zo{\"e} AE and Li, Ke and Ding, Yiliang},
  journal={Nucleic Acids Research},
  volume={52},
  number={W1},
  pages={W19--W28},
  year={2024},
  publisher={Oxford University Press}
}

@article{brazda2016palindrome,
  title={Palindrome analyser--a new web-based server for predicting and evaluating inverted repeats in nucleotide sequences},
  author={Br{\'a}zda, V{\'a}clav and Kolomazn{\'\i}k, Jan and L{\`y}sek, Ji{\v{r}}{\'\i} and H{\'a}ron{\'\i}kov{\'a}, Lucia and Coufal, Jan and {\v{S}}t'astn{\`y}, Ji{\v{r}}{\'\i}},
  journal={Biochemical and biophysical research communications},
  volume={478},
  number={4},
  pages={1739--1745},
  year={2016},
  publisher={Elsevier}
}

@article{popenda2012automated,
  title={Automated 3D structure composition for large RNAs},
  author={Popenda, Mariusz and Szachniuk, Marta and Antczak, Maciej and Purzycka, Katarzyna J and Lukasiak, Piotr and Bartol, Natalia and Blazewicz, Jacek and Adamiak, Ryszard W},
  journal={Nucleic acids research},
  volume={40},
  number={14},
  pages={e112--e112},
  year={2012},
  publisher={Oxford University Press}
}

@article{wang20193drna,
  title={3dRNA v2. 0: an updated web server for RNA 3D structure prediction},
  author={Wang, Jun and Wang, Jian and Huang, Yanzhao and Xiao, Yi},
  journal={International journal of molecular sciences},
  volume={20},
  number={17},
  pages={4116},
  year={2019},
  publisher={MDPI}
}

@article{watkins2020farfar2,
  title={FARFAR2: improved de novo rosetta prediction of complex global RNA folds},
  author={Watkins, Andrew Martin and Rangan, Ramya and Das, Rhiju},
  journal={Structure},
  volume={28},
  number={8},
  pages={963--976},
  year={2020},
  publisher={Elsevier}
}

@article{boniecki2016simrna,
  title={SimRNA: a coarse-grained method for RNA folding simulations and 3D structure prediction},
  author={Boniecki, Michal J and Lach, Grzegorz and Dawson, Wayne K and Tomala, Konrad and Lukasz, Pawel and Soltysinski, Tomasz and Rother, Kristian M and Bujnicki, Janusz M},
  journal={Nucleic acids research},
  volume={44},
  number={7},
  pages={e63--e63},
  year={2016},
  publisher={Oxford University Press}
}

@article{li2023integrating,
  title={Integrating end-to-end learning with deep geometrical potentials for ab initio RNA structure prediction},
  author={Li, Yang and Zhang, Chengxin and Feng, Chenjie and Pearce, Robin and Lydia Freddolino, P and Zhang, Yang},
  journal={Nature Communications},
  volume={14},
  number={1},
  pages={5745},
  year={2023},
  publisher={Nature Publishing Group UK London}
}

@article{wang2023trrosettarna,
  title={trRosettaRNA: automated prediction of RNA 3D structure with transformer network},
  author={Wang, Wenkai and Feng, Chenjie and Han, Renmin and Wang, Ziyi and Ye, Lisha and Du, Zongyang and Wei, Hong and Zhang, Fa and Peng, Zhenling and Yang, Jianyi},
  journal={Nature communications},
  volume={14},
  number={1},
  pages={7266},
  year={2023},
  publisher={Nature Publishing Group UK London}
}

@article{kagaya2025nufold,
  title={NuFold: end-to-end approach for RNA tertiary structure prediction with flexible nucleobase center representation},
  author={Kagaya, Yuki and Zhang, Zicong and Ibtehaz, Nabil and Wang, Xiao and Nakamura, Tsukasa and Punuru, Pranav Deep and Kihara, Daisuke},
  journal={Nature communications},
  volume={16},
  number={1},
  pages={881},
  year={2025},
  publisher={Nature Publishing Group UK London}
}

@article{tarafder2026rnabpflow,
  title={RNAbpFlow: base pair-augmented SE (3) flow matching for conditional RNA 3D structure generation},
  author={Tarafder, Sumit and Bhattacharya, Debswapna},
  journal={Nature Methods},
  pages={1--10},
  year={2026},
  publisher={Nature Publishing Group US New York}
}

@article{shen2024accurate,
  title={Accurate RNA 3D structure prediction using a language model-based deep learning approach},
  author={Shen, Tao and Hu, Zhihang and Sun, Siqi and Liu, Di and Wong, Felix and Wang, Jiuming and Chen, Jiayang and Wang, Yixuan and Hong, Liang and Xiao, Jin and others},
  journal={Nature Methods},
  volume={21},
  number={12},
  pages={2287--2298},
  year={2024},
  publisher={Nature Publishing Group US New York}
}

@article{zhang20223ddna,
  title={3dDNA: A computational method of building DNA 3D structures},
  author={Zhang, Yi and Xiong, Yiduo and Xiao, Yi},
  journal={Molecules},
  volume={27},
  number={18},
  pages={5936},
  year={2022},
  publisher={MDPI}
}

@article{wang20253d,
  title={3D structure and stability prediction of DNA with multi-way junctions in ionic solutions},
  author={Wang, Xunxun and Shi, Ya-Zhou},
  journal={PLOS Computational Biology},
  volume={21},
  number={8},
  pages={e1013346},
  year={2025},
  publisher={Public Library of Science San Francisco, CA USA}
}

@article{baek2024accurate,
  title={Accurate prediction of protein--nucleic acid complexes using RoseTTAFoldNA},
  author={Baek, Minkyung and McHugh, Ryan and Anishchenko, Ivan and Jiang, Hanlun and Baker, David and DiMaio, Frank},
  journal={Nature methods},
  volume={21},
  number={1},
  pages={117--121},
  year={2024},
  publisher={Nature Publishing Group US New York}
}

@article{abramson2024accurate,
  title={Accurate structure prediction of biomolecular interactions with AlphaFold 3},
  author={Abramson, Josh and Adler, Jonas and Dunger, Jack and Evans, Richard and Green, Tim and Pritzel, Alexander and Ronneberger, Olaf and Willmore, Lindsay and Ballard, Andrew J and Bambrick, Joshua and others},
  journal={Nature},
  volume={630},
  number={8016},
  pages={493--500},
  year={2024},
  publisher={Nature Publishing Group UK London}
}

@article{wohlwend2025boltz,
  title={Boltz-1 democratizing biomolecular interaction modeling},
  author={Wohlwend, Jeremy and Corso, Gabriele and Passaro, Saro and Getz, Noah and Reveiz, Mateo and Leidal, Ken and Swiderski, Wojtek and Atkinson, Liam and Portnoi, Tally and Chinn, Itamar and others},
  journal={BioRxiv},
  pages={2024--11},
  year={2025}
}

@article{chai2024chai,
  title={Chai-1: Decoding the molecular interactions of life},
  author={Chai Discovery team and Boitreaud, Jacques and Dent, Jack and McPartlon, Matthew and Meier, Joshua and Reis, Vinicius and Rogozhonikov, Alex and Wu, Kevin},
  journal={BioRxiv},
  pages={2024--10},
  year={2024},
  publisher={Cold Spring Harbor Laboratory}
}

@article{protenix2026protenix,
  title={Protenix-v1: Toward high-accuracy open-source biomolecular structure prediction},
  author={Protenix Team and Zhang, Yuxuan and Gong, Chengyue and Zhang, Hanyu and Ma, Wenzhi and Liu, Zhenyu and Chen, Xinshi and Guan, Jiaqi and Wang, Lan and Yang, Yanping and others},
  journal={bioRxiv},
  pages={2026--02},
  year={2026},
  publisher={Cold Spring Harbor Laboratory}
}

@article{xu2025foldbench,
  title={Benchmarking all-atom biomolecular structure prediction with FoldBench},
  author={Xu, Sheng and Feng, Qiantai and Qiao, Lifeng and Wu, Hao and Shen, Tao and Cheng, Yu and Zheng, Shuangjia and Sun, Siqi},
  journal={Nature Communications},
  volume={17},
  number={1},
  pages={442},
  year={2025},
  publisher={Nature Publishing Group UK London}
}

@article{chen2026fair,
  title={Fair splits flip the leaderboard: CHANRG reveals limited generalization in RNA secondary-structure prediction},
  author={Chen, Zhiyuan and Deng, Zhenfeng and Deng, Pan and Liao, Yue and Su, Xiu and Ye, Peng and Liu, Xihui},
  journal={arXiv preprint arXiv:2603.22330},
  year={2026}
}

@article{gong2019rna,
  title={RNA-align: quick and accurate alignment of RNA 3D structures based on size-independent TM-scoreRNA},
  author={Gong, Sha and Zhang, Chengxin and Zhang, Yang},
  journal={Bioinformatics},
  volume={35},
  number={21},
  pages={4459--4461},
  year={2019},
  publisher={Oxford University Press}
}

@article{zhu2025bpfold,
  title={Deep generalizable prediction of RNA secondary structure via base pair motif energy},
  author={Zhu, Heqin and Tang, Fenghe and Quan, Quan and Chen, Ke and Xiong, Peng and Zhou, S Kevin},
  journal={Nature Communications},
  volume={16},
  number={1},
  pages={5856},
  year={2025},
  publisher={Nature Publishing Group UK London}
}

@article{singh2019spot-rna,
  title={RNA secondary structure prediction using an ensemble of two-dimensional deep neural networks and transfer learning},
  author={Singh, Jaswinder and Hanson, Jack and Paliwal, Kuldip and Zhou, Yaoqi},
  journal={Nature communications},
  volume={10},
  number={1},
  pages={5407},
  year={2019},
  publisher={Nature Publishing Group UK London}
}

@article{penic2025rinalmo,
  title={RiNALMo: general-purpose RNA language models can generalize well on structure prediction tasks},
  author={Peni{\'c}, Rafael Josip and Vla{\v{s}}i{\'c}, Tin and Huber, Roland G and Wan, Yue and {\v{S}}iki{\'c}, Mile},
  journal={Nature Communications},
  volume={16},
  number={1},
  pages={5671},
  year={2025},
  publisher={Nature Publishing Group UK London}
}

@article{wayment2022eternafold,
  title={RNA secondary structure packages evaluated and improved by high-throughput experiments},
  author={Wayment-Steele, Hannah K and Kladwang, Wipapat and Strom, Alexandra I and Lee, Jeehyung and Treuille, Adrien and Becka, Alex and Eterna Participants and Das, Rhiju},
  journal={Nature methods},
  volume={19},
  number={10},
  pages={1234--1242},
  year={2022},
  publisher={Nature Publishing Group US New York}
}

@article{lorenz2011viennarnafold,
  title={ViennaRNA Package 2.0},
  author={Lorenz, Ronny and Bernhart, Stephan H and H{\"o}ner zu Siederdissen, Christian and Tafer, Hakim and Flamm, Christoph and Stadler, Peter F and Hofacker, Ivo L},
  journal={Algorithms for molecular biology},
  volume={6},
  number={1},
  pages={26},
  year={2011},
  publisher={Springer}
}

@article{yin2025ernie,
  title={ERNIE-RNA: an RNA language model with structure-enhanced representations},
  author={Yin, Weijie and Zhang, Zhaoyu and Zhang, Shuo and He, Liang and Zhang, Ruiyang and Jiang, Rui and Liu, Gan and Wang, Jingyi and Zhang, Xuegong and Qin, Tao and others},
  journal={Nature Communications},
  volume={16},
  number={1},
  pages={10076},
  year={2025},
  publisher={Nature Publishing Group UK London}
}

@article{zhu2025fully,
  title={A fully open structure-guided RNA foundation model for robust structural and functional inference},
  author={Zhu, Heqin and Li, Ruifeng and Chang, Ao and Chen, Haobin and Zhang, Feng and Tang, Fenghe and Ye, Tong and Li, Xin and Gu, Yunjie and Xiong, Peng and others},
  journal={bioRxiv},
  pages={2025--08},
  year={2025},
  publisher={Cold Spring Harbor Laboratory}
}

@article{sarver2008fr3d,
  title={FR3D: finding local and composite recurrent structural motifs in RNA 3D structures},
  author={Sarver, Michael and Zirbel, Craig L and Stombaugh, Jesse and Mokdad, Ali and Leontis, Neocles B},
  journal={Journal of mathematical biology},
  volume={56},
  number={1},
  pages={215--252},
  year={2008},
  publisher={Springer}
}

@article{passaro2025boltz2,
  title={Boltz-2: Towards accurate and efficient binding affinity prediction},
  author={Passaro, Saro and Corso, Gabriele and Wohlwend, Jeremy and Reveiz, Mateo and Thaler, Stephan and Somnath, Vignesh Ram and Getz, Noah and Portnoi, Tally and Roy, Julien and Stark, Hannes and others},
  journal={BioRxiv},
  year={2025}
}

@inproceedings{NCfold,
  title={NC-Bench and NCfold: A Benchmark and Closed-Loop Framework for RNA Non-Canonical Base-Pair Prediction},
  author={Zhu, Heqin and Li, Ruifeng and Chang, Ao and Li, Mingqian and Chen, Hongyang and Xiong, Peng and Zhou, Shaohua Kevin},
  year={2026},
  booktitle={The Fourteenth International Conference on Learning Representations}
}
